\documentclass[twocolumn,prd,showpacs]{revtex4}
\usepackage{amsmath}
\usepackage{amssymb}
\usepackage{graphicx}
\begin{document}
\title{Jet reconstruction in hadronic collisions by Gaussian
  filtering}

\author{Yue-Shi \surname{Lai}}
\author{Brian A.\ \surname{Cole}}
\affiliation{Columbia University, New York, NY 10027 and Nevis
  Laboratories, Irvington, NY 10533, USA}
\date{\today}

\begin{abstract}
  A new algorithm for jet finding in hadronic collisions is presented.
  The algorithm, based on a Gaussian filter in $(\eta,\phi)$, is
  specifically intended for use in heavy ion collisions and/or for
  detectors with limited acceptance. The performance of the algorithm
  is compared to two conventional algorithms, a seedless cone
  algorithm and a $k_\perp$ algorithm, for Pythia simulated di-jet
  events in $\sqrt{s} = 200\,\mathrm{GeV}$ $p + p$ collisions with
  $4\,\mathrm{GeV}/c \le \sqrt{Q^2} \le 16\,\mathrm{GeV}/c$. The
  Gaussian filter is found to perform as well as, and in some
  instances better than, the conventional algorithms.
\end{abstract}


\maketitle

\section{Introduction}

\begin{figure}
  \centerline{\includegraphics[width=3.375in]{%
      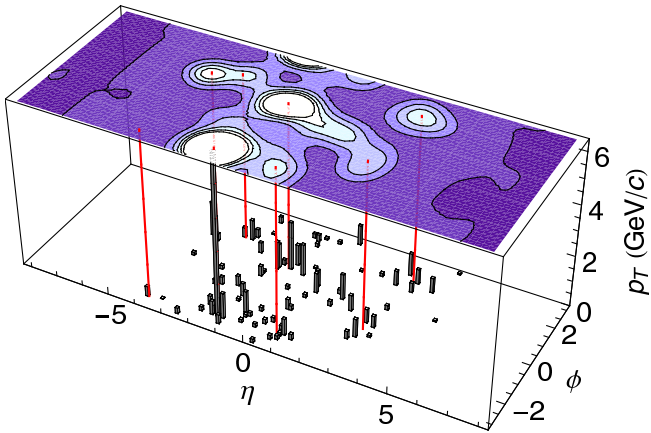}}
  \caption{A demonstration of the application of the Gaussian filter
    to a Pythia event. Final state particle $p_T$ are plotted on the
    bottom Lego plot. The result of the filter is shown with the
    contour plot on the top surface. Red connecting lines indicate
    reconstructed jet positions.}
  \label{fig:event}
\end{figure}

The observation of jet quenching in heavy ion collisions at
RHIC~\cite{Adcox:2001jp} and the problems inherent in the use of
single or di-hadron observables to quantify the energy loss of
hard-scattered partons~\cite{Adler:2005ee} together provide a strong
motivation for complete jet measurements in heavy ion collisions.
However, at RHIC energies, the underlying heavy ion event makes
application of conventional jet algorithms difficult. Commonly used
jet algorithms employ either fixed-size cones~\cite{Huth:1990mi},
augmented by split/merging procedures, or iterative clustering
(``$k_\perp$'') techniques~\cite{Catani:1993hr,Ellis:1993tq} to
determine the solid angle coverage of the jets. The kinematic
parameters of the jets are obtained by summing over the particles or
calorimeter elements within the jet with a constant
(i.e.\ angle-independent) weight. Such a flat weighting may not be
optimal in the presence of a fluctuating background because a typical
jet has most of its energy concentrated near the center of the jet and
only a small fraction of the energy in the periphery of the jet. In
contrast, an angular weighting that enhances the center of the jet
compared to the periphery would provide an improved signal to
background in the measurement of the jet energy.

Backgrounds from the underlying event in hadron--hadron collisions or
heavy ion collisions or from pile-up at high luminosity colliders can
also distort the jet finding algorithms themselves. High energy
particles from the background can prevent or modify the convergence of
the mean-shift iteration in cone algorithms or can modify the order of
combination of fragments/proto-jets in $k_\perp$ algorithms. An
algorithm that finds maxima in the the angular distribution of
fragment (transverse) energies using a weighting function or filter
that is strongly peaked has the potential to find the jet position
with less sensitivity to the presence of background. Such an algorithm
also has the advantage of reducing the impact of the limited
acceptance of certain heavy ion detectors used at RHIC and the LHC.

We describe in this paper a new algorithm for finding jets and
extracting jet kinematic quantities that uses a linear, Gaussian
filter applied in the space of pseudo-rapidity and azimuthal angle,
$(\eta,\phi)$. Jets are found as local maxima of the filter output.
Because the filter is strongly peaked, the algorithm is expected to
improve the reconstruction of jet positions and jet energies in the
presence of background and to reduce the impact of restricted aperture
on jet measurements. This paper takes the first step in exploring the
basic characteristics of the Gaussian filter jet finder by studying
its performance on Pythia simulated $p + p$ events and comparing the
performance to two conventional algorithms, the
SISCone~\cite{Salam:2007xv} seedless cone algorithm and an
implementation~\cite{Butterworth:2002xg} of the $k_\perp$ algorithm.

We note that an angular weighting of final state particles has been
used previously to define energy flow variables~\cite{Berger:2003iw},
and the algorithm described here is inspired by that work. Though we
focus here on jet finding and energy estimation, as with energy flow
variables, the filtered event shape contains more information than
just the locations and energies of reconstructed jets, A previous use
of low-pass filtering by Donati~\emph{et al.}~\cite{Donati:1983rt}
should also be recognized. However, that work only used the filter to
bootstrap other clustering algorithms, while in this work, filtering
is used for complete jet reconstruction.

\section{Algorithm}

Instead of the traditional picture of discrete final state particles
used by traditional jet reconstruction algorithms, we would like to
consider a set of particles $(p_{T,i})$ of generating the event $p_T$
density
\begin{equation}
  p_T(\eta, \phi) = \sum_{i} p_{T,i} \delta(\eta - \eta_i)
  \delta(\phi - \phi_i).
  \label{eq:final_state_particle_density}
\end{equation}
Defining the jet reconstruction over a $p_T$ density has the ability
to compensate for a non-localized background, which is useful to apply
the jet reconstruction algorithm to heavy ion collisions and
background in high luminosity hadronic colliders.

The jet reconstruction procedure can be expressed as finding the
discrete set $J$ of all jets with the transverse momentum $p_T$,
pseudo-rapidity $\eta$ and azimuth $\phi$,
\begin{equation}
  J = \{ \, (p_T, \eta, \phi) \mid p_T := \tilde{p}_T(\eta, \phi)
    \:\text{a local maximum} \, \},
    \label{eq:jet_filter}
\end{equation}
with the filtered $p_T$ density $\tilde{p}_T(\eta, \phi)$ being the
linear--circular convolution
\begin{displaymath}
  \tilde{p}_T(\eta, \phi) =
  \iint_{\mathbb{R} \times S^1} d\eta' d\phi'
  p_T(\eta', \phi') h(\eta - \eta', \phi - \phi')
\end{displaymath}
on the cylindrical $(\eta, \phi)$ topology.

\begin{figure}
  \centerline{\includegraphics[width=3.375in]{%
      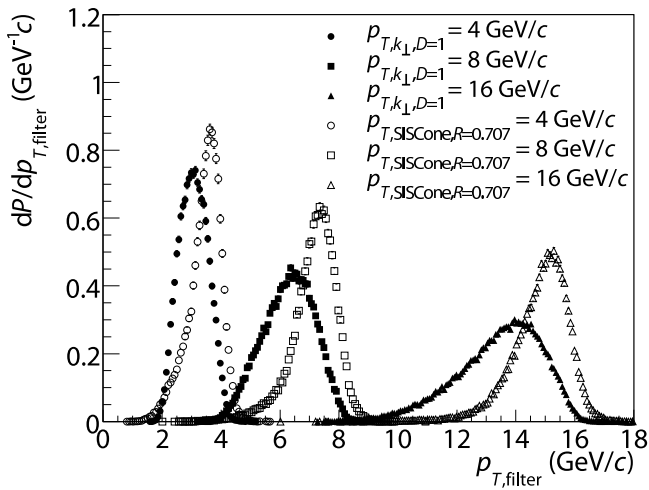}}
  \caption{Filter reconstructed Pythia jet $p_T$ distribution function
    for $p_{T,k_\perp}, p_{T,\mathrm{SISCone}} \in \{4, 8,
    16\}\,\mathrm{GeV}/c \pm 250\,\mathrm{MeV}/c$, respectively. Jets
    reconstructed by different algorithms are matched within $\Delta R
    = \sqrt{\Delta\eta^2 + \Delta\phi^2} < 0.1$.}
  \label{fig:perp_kt_siscone}
\end{figure}

When implemented solely with an iterative maximum finding,
\eqref{eq:jet_filter} would have implicitly the issue of proper
initialization, not unlike the seeding dilemma of the cone algorithm.
However, discrete digital filtering provide an efficient mean to
calculate $\tilde{p}_T(\eta, \phi)$ for a large number of sample
points. Thus one could find every possible maximum, i.e.\ in the
analogy of the cone algorithm, achieving ``seedlessnes'' by means of
sufficiently sample the entire $(\eta, \phi)$ range with a number of
seeds $N = N_\eta N_\phi \gg 2\pi\Delta\eta/R_0^2$, with $R_0 \approx
0.2$ being the characteristic separation of close jet pairs, and
$\Delta\eta = \eta_{\mathrm{max}} - \eta_{\mathrm{min}}$ the
pseudo-rapidity range (typically that of the collider experiment
calorimetry).

\begin{figure}
  \centerline{\includegraphics[width=3.375in]{%
      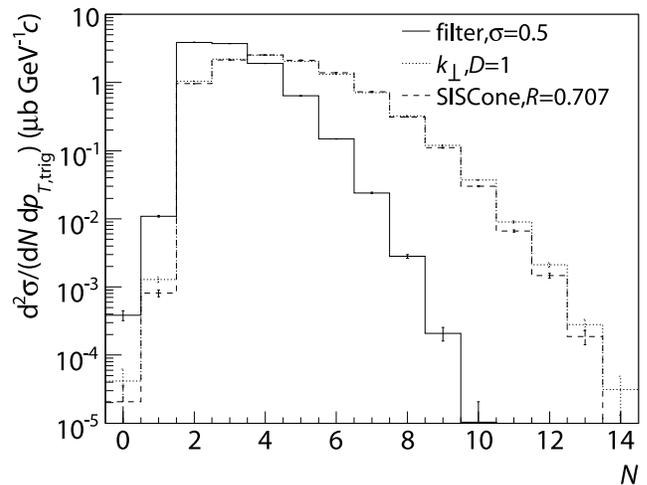}}
  \caption{Pythia jet multiplicity of jets with $p_T \ge
    2\,\mathrm{GeV}/c$, with $p_{T,\mathrm{trig}} = 8\,\mathrm{GeV}/c$
    for the filter, $k_\perp$, and SISCone algorithms}
  \label{fig:multiplicity}
\end{figure}

Therefore we implement \eqref{eq:jet_filter} by a multistage
algorithm, consisting of the following steps:

\begin{enumerate}
\item Accumulate a rectangularly binned $p_T$ density of the event.
  This can be thought as a $p_T$ histogram, or in term of the density
  distribution \eqref{eq:final_state_particle_density} and modulo a
  constant normalization, to approximate $(\eta_i, \phi_i)$ by a
  discretized $(\hat{\eta}_i, \hat{\phi}_i)$ for each $i$, with
  $\hat{\eta} = \lfloor(\eta - \eta_{\mathrm{min}})
  N_\eta/\Delta\eta\rfloor \Delta\eta/N_\eta + \eta_{\mathrm{min}}$
  (and $\lfloor \cdot \rfloor$ denoting the floor function) being the
  discretized pseudo-rapidity, and analogously for $\hat{\phi}$.
\item Apply the discrete realization of the filter on the binned
  density to obtain the initial $\tilde{p}_T(\hat{\eta}, \hat{\phi})$
  approximation for the discrete $(\hat{\eta}, \hat{\phi})$ bins or
  pixels. This filter could be either implemented in the $(\eta,
  \phi)$ position space using e.g.\ a finite impulse response (FIR) or
  infinite impulse response (IIR) version of the filter, or realized
  in the Fourier space.
\item The local maxima are localized by comparing the filtered $p_T$
  density of each discrete pixel against that of the surrounding
  pixels.
\item The set of locally maximal pixel centers $(\hat{\eta},
  \hat{\phi})$ is used to initialize a suitable local optimization
  algorithm, operating on the unbinned $p_T$ density
  \eqref{eq:final_state_particle_density}, to obtain the true $(\eta,
  \phi)$ positions.
\end{enumerate}

\begin{figure*}
  \centerline{\includegraphics[width=6.75in]{%
      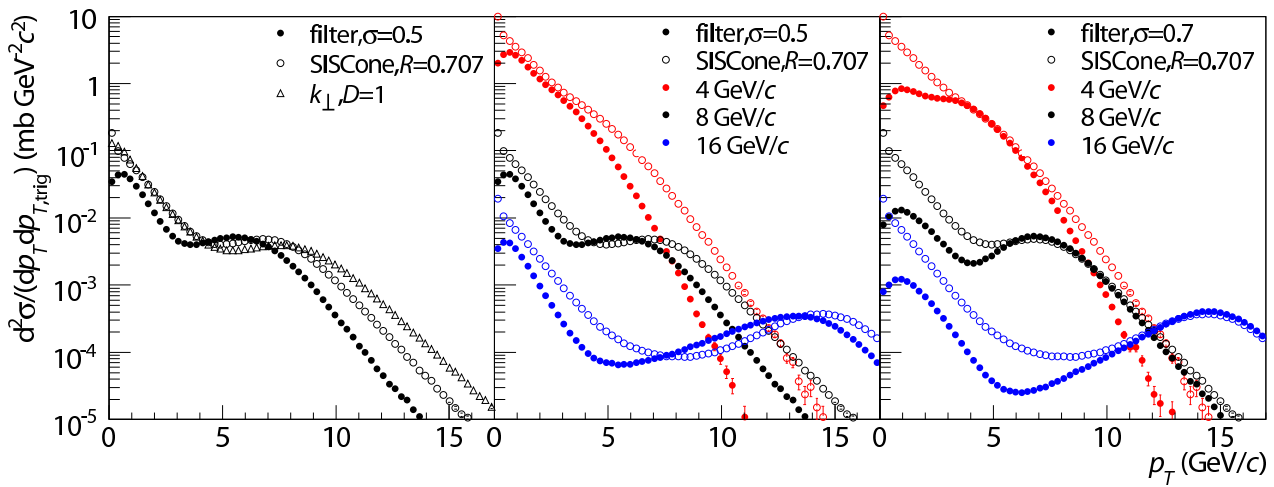}}
  \caption{Triggered Pythia jet spectrum, left with
    $p_{T,\mathrm{trig}} = 8\,\mathrm{GeV}/c$, comparing different
    algorithm and size $\sigma$ selection, and to the middle comparing
    filter $\sigma = 0.5$ and $\sigma = 0.7$, respectively, against
    SISCone $R = 2^{-1/2} \approx 0.707$ and for different
    $p_{T,\mathrm{trig}} \in \{4, 8, 16\}\,\mathrm{GeV}/c$}
  \label{fig:perp_triggered}
\end{figure*}

A jet definition like \eqref{eq:jet_filter} is inherently collinearly
and infrared safe, as $\tilde{p}_T(\eta, \phi)$ is a non-iteratively
defined quantity not involving any thresholds or cutoffs, and
therefore like event shape variables such as thrust, is neither
sensitive to infinitely soft radiation, nor collinear splitting. When
implemented properly, the maximum finding only depends on the
collinearly and infrared safe $\tilde{p}_T(\eta, \phi)$.

For the filtering kernel, we would like to propose a bivariate
Gaussian distribution function with the normalization $h(0, 0) = 1$,
\begin{equation}
  h(\eta, \phi) = \exp\left[ -(\eta^2 + \phi_\mathrm{ar}^2) /
    (2\sigma^2) \right],
  \label{eq:gaussian_kernel}
\end{equation}
with $\phi_\mathrm{ar} := 2\pi \lceil (\phi + \pi) / (2\pi) \rceil -
\pi$ (and $\lceil \cdot \rceil$ denoting the ceiling function) being
the angularly reduced azimuth. The parameter $\sigma$ determines the
radial scale of the jet reconstruction, analogous to the fixed-cone
radius $R$. Figure~\ref{fig:event} illustrates the Gaussian filter
applied to a Pythia event, and a comparison of the filter
reconstructed jet $p_T$ distribution functions for jets that the
$k_\perp$ algorithm and SISCone reconstruct to $p_T = 4$, $8$, and
$16\,\mathrm{GeV}/c$ is shown in Figure~\ref{fig:perp_kt_siscone}.

The mean shift iteration (such as the cone algorithm) is closely
related to a bound maximization with respect to the convolution with a
``shadow function''~\cite{Cheng:1995ms,Fashing:2005ms}. This has the
consequence that for the pure jet direction finding, the filtering
based algorithm with subsequent maximization presented here has an
exact correspondence in the picture of a weighted mean shift
iteration, and therefore can be regarded as a generalized form of the
cone algorithm. The correspondence also leads naturally to the
definition of jets as the local maxima in \eqref{eq:jet_filter}. For
the cone algorithm, the shadow function kernel is $h(\eta, \phi) =
\max[0, 1 - (\eta^2 + \phi^2) / R^2]$. Comparing this with the
functional form of \eqref{eq:gaussian_kernel} suggest for the limit of
narrowly focused jet, the angular behavior of the Gaussian filter and
the cone algorithm (without split/merge) would correspond with the
choice $\sqrt{2} \sigma = R$.

There are several reasons to prefer a Gaussian density distribution
instead of e.g.\ the parabolic shadow function implied by the cone
algorithm. A Gaussian density does not possess the undesirable feature
of maximum creation (or in the language of a cone iteration, stable
midpoint axes). And when a slow varying background is present, the
foreground to background ratio is accounted for both in the
reconstructed jet energy and the directional jet finding, since the
Gaussian weighting is present in both $\tilde{p}_T(\eta, \phi)$ and
its gradient. As a rapidly decreasing test function, the Gaussian
density possesses the mathematical property necessary to regulate the
singular $(\eta, \phi)$ directional distribution of an outgoing
high-$p_T$ fragment in \eqref{eq:final_state_particle_density}.

Gaussian filtering has been used extensively in fields such as
computer vision, and several fast approximations are known beside the
direct and Fourier space convolution. We use a partial fraction
expanded form~\cite{Hale:2006rg} of the IIR approximation originally
described by Young \& van
Vliet~\cite{Young:1995ri,Vliet:1998rg,Young:2002rg}, which compared to
a fast Fourier transform (FFT) realization requires no padding along
the pseudo-rapidity axis, but at least two filtering passes to
simulate a circular response in azimuth. It should be noted that
multi-pass filtering does not exactly reproduce in the the angularly
truncated Gaussian distribution as in \eqref{eq:gaussian_kernel}, but
rather the sum of its circular frequency components, and the filtering
kernel therefore in fact takes the form
\begin{displaymath}
  h(\eta, \phi) = (2\pi)^{-1/2} \lvert\sigma\rvert
  e^{-\eta^2/(2\sigma^2)} \vartheta_3\bigl( \phi/2, e^{-\sigma^2/2}
  \bigr),
\end{displaymath}
with $\vartheta_3$ being the Jacobi theta
function~\cite{Abramowitz:1972et}. The difference is however
negligible for all meaningful choices of $\sigma \ll \pi$, especially
at the initial, approximate stage of filtering.

As input to the filter we consider final state particles within an
acceptance of $\lvert\eta\rvert < 8$. The discrete filtering operates
on a $N_\eta = 640, N_\phi = 256$ grid. For the continuous
maximization, we use the Newton optimization algorithm with
analytically calculated gradient and Hessian, and for robustness,
modified to handle of indefinite Hessian by spectral decomposition.

\section{Simulation and event selection}

In order to assess the performance of jet reconstruction by filtering,
we compared the reconstructed jets against the $k_\perp$ and cone
algorithms at the ideal detector level, by using event generator truth
particles from Pythia 6.4.16~\cite{Sjostrand:2006za}.

For all comparisons we present here, we consider the $p + p$ collision
system at $\sqrt{s} = 200\,\mathrm{GeV}$, which as of Run-8 accounts
for approximately $170\,\mathrm{pb}^{-1}$ of the hadronic RHIC
operating mode. The $k_\perp$ algorithm used is the
KtJet~\cite{Butterworth:2002xg} implementation in the inclusive
clustering mode, with the inclusive stopping parameter $D = 1$ and the
longitudinally invariant QCD distance scheme~\cite{Catani:1993hr}. We
further compare against SISCone~\cite{Salam:2007xv}, a variant of the
cone algorithm, with the overlap threshold parameter $f = 0.5$ and the
cone radius $R = 2^{-1/2} \approx 0.707$, close to the typically used
$R = 0.7$ and in asymptotic angular correspondence to the Gaussian
filter with $\sigma = 0.5$ for narrow jets. Each of the comparison
presented here are based on $10^6$ (triggered) Pythia events.

\begin{figure}
  \centerline{\includegraphics[width=3.375in]{%
      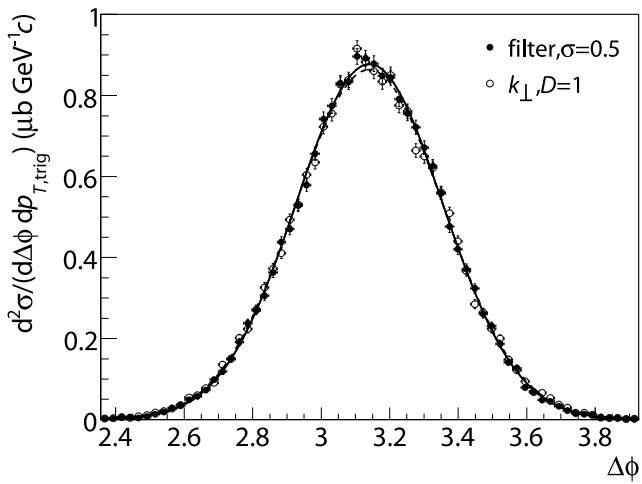}}
  \caption{Pythia di-jet event jet opening angle with
    $p_{T,\mathrm{trig}} = 8\,\mathrm{GeV}/c$ for the filter and
    $k_\perp$ algorithms}
  \label{fig:balance_angle}
\end{figure}

\begin{figure}
  \centerline{\includegraphics[width=3.375in]{%
      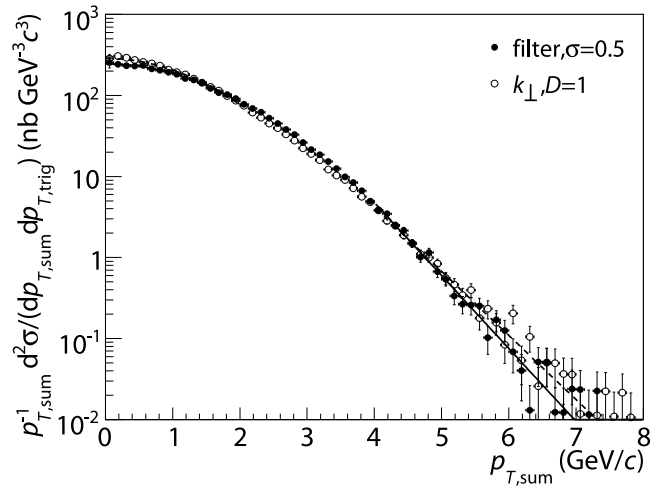}}
  \caption{Pythia di-jet event $p_T$ sum with $p_{T,\mathrm{trig}} =
    8\,\mathrm{GeV}/c$ for the filter and $k_\perp$ algorithms, with
    the $k_\perp$ convolved by the filter to $k_\perp$ energy scale
    ratio. A Gaussian and exponential fit for the low and high
    inbalance, respectively, are shown.}
  \label{fig:balance_perp}
\end{figure}

\begin{figure}[b]
  \centerline{\includegraphics[width=3.375in]{%
      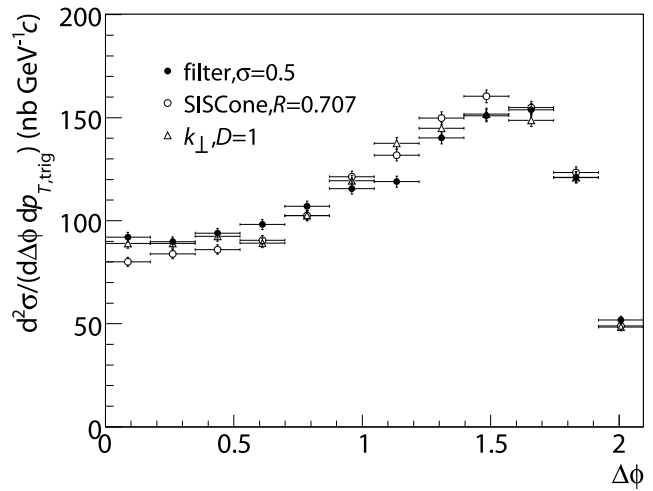}}
  \caption{Pythia three jet event jet opening angle with
    $p_{T,\mathrm{trig}} = 8\,\mathrm{GeV}/c$ for the filter and
    $k_\perp$ algorithms}
  \label{fig:threejet_angle}
\end{figure}

Two form of triggering were used. For the event multiplicity, di-jet
and three jet balance (Figures~\ref{fig:multiplicity},
\ref{fig:balance_angle}, \ref{fig:balance_perp},
\ref{fig:threejet_angle}), we are comparing two algorithms at a
certain scale of hard scattering. Here we triggered on the Pythia hard
scattering $\hat{p}_{T,2\rightarrow 2}$ with a centered,
$500\,\mathrm{MeV}$ wide window, i.e.\ with $p_{T,\mathrm{trig}} -
0.25\,\mathrm{GeV}/c \le \hat{p}_{T,2\rightarrow 2} \le
p_{T,\mathrm{trig}} + 0.25\,\mathrm{GeV}/c$, and obtained the
differential cross sections per unit trigger window at the triggered
$p_{T,\mathrm{trig}}$. The same applies to the triggered $p_T$
spectrum that illustrates the sensitivity to soft QCD background
(Figure~\ref{fig:perp_triggered}). For the comparison of filter
reconstructed $p_T$ against the reconstructed $p_T$ of the $k_\perp$
and cone algorithms (Figure~\ref{fig:perp_kt_siscone}), we used a
large trigger window with $\hat{p}_{T,2\rightarrow 2} \ge
p_{T,\mathrm{trig}} - 2.25\,\mathrm{GeV}/c$ to avoid biasing the jet,
and the results are shown as normalized probability density functions.

To test the di-jet balance of the filtering algorithm, we arrange the
reconstructed jets in descending $p_T$, i.e.\ $p_{T,1} \ge p_{T,2} \ge
\dots \ge p_{T,N}$. Events with a prominent di-jet structure can be
selected by requiring $p_{T,3} < g p_{T,2}$, with $g < 1$ as the
relative $p_T$ gap, which we chose to be $\frac{1}{4}$. This approach
avoids sensitivity to the individual energy scales of the algorithm.
In the pairwise comparision, the selection is applied on both jet
algorithms in the logical ``and'' sense to ensure symmetric event
selection (i.e. both algorithms agree that a clear dijet structure is
present).

An analogous selection for the three jet events can be made by
requiring $p_{T,4} < g p_{T,3}$ (with the same $g = \frac{1}{4}$
choice as with the di-jet event selection). If the pairwise angle
among the three leading jets are $\phi_{12}, \phi_{23}, \phi_{31}$,
the resulting three jet opening angle is defined as $\Delta\phi :=
\min(\phi_{12}, \phi_{23}, \phi_{31})$. Since three algorithms are
compared in Figure~\ref{fig:threejet_angle}, we used the logical
``or'' sense (i.e. only one algorithm detecting a clear three jet
structure is sufficient).

The inbalance expressed by the three momentum $p_T$ sum scales with
the $p_T$ scale the jet algorithm reconstructs to. Therefore the
comparison is made against the $k_\perp$ algorithm with the momentum
sum convoluted by the $p_{T,\mathrm{filter}}/p_{T,k_\perp}$
distribution function, i.e.\ if one would ``simulate'' the filter
behavior by apply the distribution function event by event to the
$k_\perp$ algorithm.

While not explicitly shown, we also checked our results against
\textsc{herwig} 5.6.10~\cite{Corcella:2000bw}, and found no algorithm
specific, dissimilar behaviors, beside differences among the event
generators that are reproduced by all algorithms.

\section{Discussion}

The prominent feature of jet reconstruction by filtering is the effect
of the angular weight. It has the effect that the reconstructed jet
$p_T$ is shifted for wide angle fragmentation, while becoming less
noticeable for sharply focused, high $p_T$ jets
(Figure~\ref{fig:perp_kt_siscone}). This accounts for the performance
of the filter algorithm in the rejection of the soft background, which
usually only contributes significantly to the event $p_T$ when summed
over large angles. This inherent discrimination against the background
provide us with a good start position to apply the formalism presented
here to stronger background levels, including those usually found in
heavy ion collisions. And while this effect complicates the
determination of jet $p_T$, the parameter dependent shift of jet $p_T$
and the subsequent need for calibration is present with all jet
reconstruction algorithms.

To show the insensitivity to the background, we calculated both the
jet multiplicity and jet $p_T$ specturm at $8\,\mathrm{GeV}/c$ initial
hard scattering. In term of the jet multiplicity
(Figure~\ref{fig:multiplicity}), the filtering reproduces more
accurately an initial hard scattering like event shape, which is
dominated by two and three jet events.

The $p_T$ spectra (Figure~\ref{fig:perp_triggered}) shows a two
component mixture, consisting of a high jet cross section, low $p_T$
exponential distribution resulting from the soft QCD background, and a
peak from the triggered, initial hard scattering. Here both the
$k_\perp$ algorithm and SISCone in fact produce a large amount of
``jets'' from background particles, thus creating the large jet
multiplicity visible in Figure~\ref{fig:multiplicity}, while the
Gaussian filter suppresses the background approximately by a factor of
2 at an initial hard scattering of $8\,\mathrm{GeV}/c$, thus producing
a significantly more isolated and prominent peak for the trigger. This
effect is consistent over a range of filter sizes and
$p_{T,\mathrm{trig}}$, and becomes even stronger with increasing
filter size or at lower $p_{T,\mathrm{trig}}$. While SISCone
e.g.\ fails to resolve the hard scattering cross section peak at
$4\,\mathrm{GeV}/c$ from the background for any choice of $R <
\frac{\pi}{2}$, this is possible with the filter algorithm and $\sigma
\gtrsim 0.8$, albeit not at $\sigma = 0.5$.

In di-jet events, the Gaussian filter reproduces the angular behavior
that previous jet reconstruction algorithms exhibit
(Figures~\ref{fig:balance_angle}), and a slight difference in the
three momentum sum is visible, with the filter having a slightly
weaker exponential tail (Figure~\ref{fig:balance_perp}). The
similarities are expected, since with adequate jet reconstruction
performance, these variables should be dominated by the physical
processes rather than the jet definitions. There is also a slight
difference in the three jet opening angle when compared to the
$k_\perp$ algorithm (\ref{fig:threejet_angle}), and difference to the
$k_\perp$ algorithm is smaller than the $k_\perp$ algorithm to
SISCone.

\section{Conclusion}

In this paper, we presented a new algorithm for jet reconstruction by
filtering, that beside essential pQCD properties of collinear and
infrared safety, provides robustness against soft background. Applied
on event generator final state particles, we showed that its
performance in key areas matches the $k_\perp$ algorithm and SISCone,
such as for the di-jet balance and three jet opening angle, which are
given by the physical jet production and jet fragmentation and not
significantly modified by the jet definition. While the $p_T$ scale
differs from the reconstructed scale of the $k_\perp$ and cone
algorithms, we note that an energy calibration is usually required for
all algorithms for the extraction of the accurate parton energy. We
also demonstrated, in this paper for $p + p$ collisions, its superior
ability in discriminating jets from the background, thus rejecting the
associated, spurious jets contaminating events at low $p_T$. We hope
to apply the algorithm presented here to most of the collision systems
available at RHIC and LHC, thus providing an unified jet definition
suitable across a large range of species, collision energies and jet
$p_T$ that so far was not attainable with the traditional jet
reconstruction algorithms.

\bibliography{bib/jetalgpp}

\end{document}